
\input harvmac

\input graphicx

%
%
\ifx\includegraphics\UnDeFiNeD\message{(NO graphicx.tex, FIGURES WILL BE
IGNORED)}
\def\figin#1{\vskip2in}
\else\message{(FIGURES WILL BE INCLUDED)}\def\figin#1{#1}
\fi
\def\Fig#1{Fig.~\the\figno\xdef#1{Fig.~\the\figno}\global\advance\figno
 by1}
%
%
%
%
\def\Ifig#1#2#3#4{
\goodbreak\midinsert
\figin{\centerline{
\includegraphics[width=#4truein]{#3}}}
\narrower\narrower\noindent{\footnotefont
{\bf #1:}  #2\par}
\endinsert
}
%

\noblackbox

\def\physrev#1#2#3{Phys. Rev. D {#1} (#2) #3}

\def\tilde{\widetilde}

\def\Qbar{\overline{Q}}
\def\Pbar{\overline{P}}


\def\drawbox#1#2{\hrule height#2pt
        \hbox{\vrule width#2pt height#1pt \kern#1pt
               \vrule width#2pt}
               \hrule height#2pt}


\def\Fund#1#2{\vcenter{\vbox{\drawbox{#1}{#2}}}}
\def\Asym#1#2{\vcenter{\vbox{\drawbox{#1}{#2}
              \kern-#2pt       
              \drawbox{#1}{#2}}}}
\def\Asymt#1#2{\vcenter{\vbox{\drawbox{#1}{#2}
              \kern-#2pt       
              \drawbox{#1}{#2}
              \kern-#2pt       
              \drawbox{#1}{#2}}}}

\def\fund{\Fund{6.5}{0.4}}

\baselineskip=12pt

\hfill{RUNHETC-07-08}

\Title{}
{\vbox{\centerline{Meta-Stable Dynamical Supersymmetry Breaking }
\vglue 0.175in
\centerline{Near Points of Enhanced Symmetry}} }

\bigskip
\centerline{{\tenbf Rouven Essig, Kuver Sinha, Gonzalo Torroba}}
\vglue .5cm
\centerline{New High Energy Theory Center }
\centerline{Physics Department}
\centerline{Rutgers University}
\centerline{Piscataway, NJ 08854}

\vglue .2in

\bigskip

\noindent
We show that metastable supersymmetry breaking is generic 
near certain enhanced symmetry points of gauge theory moduli spaces.
Our model consists of two sectors coupled by a singlet and combines 
dynamical supersymmetry breaking with an O'Raifeartaigh mechanism in 
terms of confined variables.
All relevant mass parameters, including the supersymmetry
breaking scale, are generated dynamically.
The metastable vacua appear as a result of a balance between
non-perturbative and perturbative quantum effects along a pseudo-runaway direction.

\vglue .8in
\noindent{---------------------------------------}

\noindent{rouven, kuver, torrobag @physics.rutgers.edu}

\Date{}

\lref\mike{F.~Denef and M.~R.~Douglas,
  ``Distributions of nonsupersymmetric flux vacua,''
  JHEP {\bf 0503}, 061 (2005)
  [arXiv:hep-th/0411183];~ F.~Denef and M.~R.~Douglas,
  ``Distributions of flux vacua,''
  JHEP {\bf 0405}, 072 (2004)
  [arXiv:hep-th/0404116];
  ~M.~R.~Douglas,
  ``Statistical analysis of the supersymmetry breaking scale,''
  [arXiv:hep-th/0405279].
}

\lref\iss{K.~Intriligator, N.~Seiberg and D.~Shih,
  ``Dynamical SUSY breaking in meta-stable vacua,''
  JHEP {\bf 0604}, 021 (2006)
  [arXiv:hep-th/0602239].
  }

\lref\scott{S.~D.~Thomas,
  ``Recent developments in dynamical supersymmetry breaking,''
  [arXiv: hep-th/9801007].}

\lref\shih{D.~Shih,
  ``Spontaneous R-symmetry breaking in O'Raifeartaigh models,''
  [arXiv:hep-th/0703196].}

\lref\review{K.~Intriligator and N.~Seiberg,
  ``Lectures on Supersymmetry Breaking,''
  [arXiv:hep-ph/0702069].}

\lref\Korneel{K.~v.~d.~Broek,
  ``Vscape V1.1.0: An interactive tool for metastable vacua,''
  arXiv: 0705.2019 [hep-ph].}

\lref\three{
  I.~Affleck, M.~Dine and N.~Seiberg,
  ``Dynamical Supersymmetry Breaking In Four-Dimensions And Its
  Phenomenological Implications,''
  Nucl.\ Phys.\  B {\bf 256}, 557 (1985);
}

\lref\ADS{
  ~I.~Affleck, M.~Dine and N.~Seiberg,
  ``Exponential Hierarchy From Dynamical Supersymmetry Breaking,''
  Phys.\ Lett.\  B {\bf 140}, 59 (1984);
  ~I.~Affleck, M.~Dine and N.~Seiberg,
  ``Calculable Nonperturbative Supersymmetry Breaking,''
  Phys.\ Rev.\ Lett.\  {\bf 52}, 1677 (1984);
  ~I.~Affleck, M.~Dine and N.~Seiberg,
  ``Dynamical Supersymmetry Breaking In Chiral Theories,''
  Phys.\ Lett.\  B {\bf 137}, 187 (1984);
  ~I.~Affleck, M.~Dine and N.~Seiberg,
  ``Dynamical Supersymmetry Breaking In Supersymmetric QCD,''
  Nucl.\ Phys.\  B {\bf 241}, 493 (1984);
  ~I.~Affleck, M.~Dine and N.~Seiberg,
  ``Supersymmetry Breaking By Instantons,''
  Phys.\ Rev.\ Lett.\  {\bf 51}, 1026 (1983).
}

\lref\issa{
  K.~Intriligator, N.~Seiberg and D.~Shih,
  ``Supersymmetry Breaking, R-Symmetry Breaking and Metastable Vacua,''
  arXiv:hep-th/0703281.
}

\lref\coleman{S.R. Coleman, ``The Fate of the False Vacuum. 1. Semiclassical Theory,'' \physrev{15}{2929}{1977}, Erratum-ibid. D 16 (1248) 1977.}

\lref\duncan{M.J. Duncan and L.G. Jensen, ``Exact tunneling solutions in scalar field theory,'' Phys. Lett. B {\bf 291} (1995) 109.}

\lref\WittenDSB{
  E.~Witten,
  ``Dynamical Breaking Of Supersymmetry,''
  Nucl.\ Phys.\  B {\bf 188}, 513 (1981).
}

\lref\Wittenindex{
  E.~Witten,
  ``Constraints On Supersymmetry Breaking,''
  Nucl.\ Phys.\  B {\bf 202}, 253 (1982).
}

\lref\SeibergNelson{
  A.~E.~Nelson and N.~Seiberg,
  ``R symmetry breaking versus supersymmetry breaking,''
  Nucl.\ Phys.\  B {\bf 416}, 46 (1994)
  [arXiv:hep-ph/9309299].
}

\lref\CW{
  S.~R.~Coleman and E.~Weinberg,
  ``Radiative Corrections As The Origin Of Spontaneous Symmetry Breaking,''
  Phys.\ Rev.\  D {\bf 7}, 1888 (1973).
}

\lref\oraf{
  L.~O'Raifeartaigh,
  ``Spontaneous Symmetry Breaking For Chiral Scalar Superfields,''
  Nucl.\ Phys.\  B {\bf 96}, 331 (1975).
}
\lref\brane{A.~Giveon and D.~Kutasov,
  ``Gauge symmetry and supersymmetry breaking from intersecting branes,''
  [arXiv:hep-th/0703135];~
M.~Aganagic, C.~Beem, J.~Seo and C.~Vafa,
  ``Geometrically induced metastability and holography,''
  [arXiv: hep-th / 0610249]; ~
H.~Ooguri  and Y.~Ookouchi,
  ``Meta-stable supersymmetry breaking vacua on intersecting branes,''
  Phys.\ Lett.\  B {\bf 641}, 323 (2006)
  [arXiv:hep-th/0607183]; ~S.~Franco, I.~Garcia-Etxebarria and A.~M.~Uranga,
  ``Non-supersymmetric meta-stable vacua from brane configurations,''
  JHEP {\bf 0701}, 085 (2007)
  [arXiv:hep-th/0607218].
}
\lref\franco{
R.~Argurio, M.~Bertolini, S.~Franco and S.~Kachru,
  ``Gauge/gravity duality and meta-stable dynamical supersymmetry breaking,''
  JHEP {\bf 0701}, 083 (2007)
  [arXiv:hep-th/0610212];~R.~Argurio, M.~Bertolini, S.~Franco and S.~Kachru,
  ``Metastable vacua and D-branes at the conifold,''
  JHEP {\bf 0706}, 017 (2007)
  [arXiv:hep-th/0703236].
}

\lref\dinea{
  M.~Dine and Z.~Sun,
  ``R symmetries in the landscape,''
  JHEP {\bf 0601}, 129 (2006)
  [arXiv:hep-th/0506246].
}

\lref\warp{S.~B.~Giddings, S.~Kachru and J.~Polchinski,
  ``Hierarchies from fluxes in string compactifications,''
  Phys.\ Rev.\  D {\bf 66}, 106006 (2002)
  [arXiv:hep-th/0105097];~ S.~Kachru, R.~Kallosh, A.~Linde and S.~P.~Trivedi,
  ``De Sitter vacua in string theory,''
  Phys.\ Rev.\  D {\bf 68}, 046005 (2003)
  [arXiv:hep-th/0301240]; ~ M.~R.~Douglas, J.~Shelton and G.~Torroba,
  ``Warping and supersymmetry breaking,''
  arXiv:0704.4001 [hep-th];~ C.~P.~Burgess, P.~G.~Camara, S.~P.~de Alwis, S.~B.~Giddings, A.~Maharana, F.~Quevedo and K.~Suruliz,
  ``Warped supersymmetry breaking,''
  [arXiv:hep-th/0610255];~ S.~Kachru, J.~Pearson and H.~L.~Verlinde,
  ``Brane/flux annihilation and the string dual of a non-supersymmetric  field
  theory,''
  JHEP {\bf 0206}, 021 (2002)
  [arXiv:hep-th/0112197].
}

\lref\cy{D.~E.~Diaconescu, A.~Garcia-Raboso and K.~Sinha,
  ``A D-brane landscape on Calabi-Yau manifolds,''
  JHEP {\bf 0606}, 058 (2006)
  [arXiv:hep-th/0602138];
  D.~E.~Diaconescu, A.~Garcia-Raboso, R.~L.~Karp and K.~Sinha,
  ``D-brane superpotentials in Calabi-Yau orientifolds (projection),''
  [arXiv:hep-th/0606180];~
  ~S.~Franco and A.~M.~.~Uranga,
  ``Dynamical SUSY breaking at meta-stable minima from D-branes at obstructed
  geometries,''
  JHEP {\bf 0606}, 031 (2006)
  [arXiv:hep-th/0604136]; ~D.~E.~Diaconescu, R.~Donagi and B.~Florea,
  ``Metastable quivers in string compactifications,''
  [arXiv:hep-th/0701104].
}

\lref\twoa{O.~DeWolfe, A.~Giryavets, S.~Kachru and W.~Taylor,
  ``Type IIA moduli stabilization,''
  JHEP {\bf 0507}, 066 (2005)
  [arXiv:hep-th/0505160];
  ~T.~Banks and K.~van den Broek,
  ``Massive IIA flux compactifications and U-dualities,''
  JHEP {\bf 0703}, 068 (2007)
  [arXiv:hep-th/0611185];
  ~I.~Bena, E.~Gorbatov, S.~Hellerman, N.~Seiberg and D.~Shih,
  ``A note on (meta)stable brane configurations in MQCD,''
  JHEP {\bf 0611}, 088 (2006)
  [arXiv:hep-th/0608157].
}

\lref\Banks{
  T.~Banks,
  ``Remodeling the pentagon after the events of 2/23/06,''
  arXiv:hep-ph/0606313;
  ~T.~Banks,
  ``Cosmological supersymmetry breaking and the power of the pentagon: A  model
  of low energy particle physics,''
  arXiv:hep-ph/0510159;
  ~T.~Banks, S.~Echols and J.~L.~Jones,
  ``Baryogenesis, dark matter and the pentagon,''
  JHEP {\bf 0611}, 046 (2006)
  [arXiv:hep-ph/0608104].
}

\lref\mediation{
  O.~Aharony and N.~Seiberg,
  ``Naturalized and simplified gauge mediation,''
  JHEP {\bf 0702}, 054 (2007)
  [arXiv:hep-ph/0612308]; ~
M.~Dine and J.~Mason,
``Gauge mediation in metastable vacua,''
  [arXiv:hep-ph/0611312];~
R.~Kitano, H.~Ooguri and Y.~Ookouchi,
  ``Direct mediation of meta-stable supersymmetry breaking,''
  Phys.\ Rev.\  D {\bf 75}, 045022 (2007)
  [arXiv:hep-ph/0612139]; ~ D.~E.~Diaconescu, B.~Florea, S.~Kachru and P.~Svrcek,
  ``Gauge - mediated supersymmetry breaking in string compactifications,''
  JHEP {\bf 0602}, 020 (2006)
  [arXiv:hep-th/0512170].
}

\lref\retro{M.~Dine, J.~L.~Feng and E.~Silverstein,
  ``Retrofitting O'Raifeartaigh models with dynamical scales,''
  Phys.\ Rev.\  D {\bf 74}, 095012 (2006)
  [arXiv:hep-th/0608159].
}

\lref\felix{
  F.~Brummer,
  ``A natural renormalizable model of metastable SUSY breaking,''
  arXiv:0705.2153 [hep-ph].
}
\lref\MalyshevYB{
  D.~Malyshev,
  ``Del Pezzo singularities and SUSY breaking,''
  arXiv:0705.3281 [hep-th].
}

\lref\sem{
K.~A.~Intriligator and N.~Seiberg,
  ``Lectures on supersymmetric gauge theories and electric-magnetic  duality,''
  Nucl.\ Phys.\ Proc.\ Suppl.\  {\bf 45BC}, 1 (1996)
  [arXiv:hep-th/9509066].
}

\baselineskip=16pt


\newsec{Introduction}

The idea that our universe may be in a long-lived 
metastable state in which supersymmetry is broken
has recently led to an increased interest in the construction 
of models of supersymmetry breaking.  
This has opened many new possibilities in constructing field theory and 
string theory models.

On the field theoretic side, the work of Intriligator, Seiberg and Shih (ISS) $\iss$ constructed 
calculable metastable vacua using Seiberg duality. 
This motivated related field theory constructions, involving gauge 
mediation $\mediation$, generalized O'Raifeartaigh models $\shih$, 
retrofitting $\retro$, applications to particle physics $\Banks$ etc. 
Similar developments have been seen in string theory based on a number 
of different tools, such as intersecting or wrapping branes $\brane$, flux
compactifications $\warp$, Calabi-Yau's with particular geometric properties 
$\cy$, IIa/M-theory configurations $\twoa$ and others.
Statistical analyses of the supersymmetry breaking scale on the landscape 
of effective field theories were done, for instance, in $\mike$.

The ISS model consists of SQCD in the free magnetic range, and metastable vacua appear after taking into 
account one-loop corrections that lift the pseudo-moduli. Their work
suggests that nonsupersymmetric vacua are rather generic, if one requires them to be only local, rather than global,
minima of the potential. The construction
still contained relevant couplings in the form of masses for the quarks though, and the search
for models with all the relevant parameters generated dynamically has proven difficult;
see $\franco$, $\felix$, $\MalyshevYB$,  for recent work in this direction. 

One lesson from ISS is that certain properties of moduli space can hint at the existence of 
metastable vacua.  In their case, it was the existence of supersymmetric vacua coming in from infinity 
that signaled an approximate R-symmetry.  Here we will point out that one should also look for another 
feature, namely, enhanced symmetry points, which are defined by the appearance of massless particles.
We claim that if the moduli space has certain coincident enhanced symmetry points, metastable
vacua with all the relevant couplings arising by dimensional transmutation may be obtained.

Let us motivate this claim.  In order to generate relevant couplings dynamically, 
a gauge sector is required, which gives nonperturbative contributions to the superpotential.  
However, in general this leads to a runaway behavior.  We will show that starting with two 
gauge sectors, the runaway may now be stabilized by one loop effects from the additional gauge 
sector, but only around enhanced symmetry points where quantum corrections are large enough.
Such runaways which are stabilized by perturbative quantum corrections will be called `pseudo-runaways'. 
Surprisingly, the gauge theories where this occurs turn out to be generic.

The model considered here consists of two SQCD sectors, each with independent 
rank and number of flavors, coupled by a singlet.  It involves 
only marginal operators with all scales generated dynamically.
At the origin of moduli space, the singlet vanishes and the quarks of 
both sectors become massless simultaneously.  
There are thus two coincident enhanced symmetry points at the origin.
While one of the SQCD sectors is in the electric range and produces a 
runaway, the other has a magnetic dual description as 
an O'Raifeartaigh-like model.  Near the enhanced symmetry point, the 
Coleman-Weinberg corrections stabilize the nonperturbative instability producing 
a long-lived metastable vacuum.
A feature of our model is that it may be possible to gauge parts of 
its large global symmetry to obtain renormalizable, natural models of direct gauge mediated
supersymmetry breaking with a singlet. R-symmetry is broken both
spontaneously and explicitly in our model.

The plan of the paper is as follows. In Section 2, our model is
introduced and its supersymmetric vacua are studied. In Section 3, we
analyze in detail the non-supersymmetric vacua and argue that they are parametrically 
long-lived.  In Section 4, we give a detailed analysis of the particle spectrum
and the R-symmetry properties.  In Section 5, we argue that such metastable vacua 
may be generic near points of enhanced symmetry in the landscape of effective 
field theories.  In Section 6, we give our conclusions.


\newsec{The Model and its Supersymmetric Vacua}

We consider models with two supersymmetric QCD (SQCD) sectors characterized by $(N_c, N_f, \Lambda)$ and 
$(N_c',N_f',\Lambda')$, respectively, that are coupled to the same singlet field $\Phi$.  
The field $\Phi$ provides the mass of the quarks in both sectors.  
In Section 2.1, the general properties of such models will be discussed and their global symmetries analyzed.
In Section 2.2, we analyze the supersymmetric vacua.  
Section 2.3 will discuss for which range of the parameters $(N_c, N_f, \Lambda)$ and $(N_c',N_f',\Lambda')$ 
metastable vacua will be shown to exist.  The upshot will be that one sector has to be taken in 
the electric range and the other sector in the free magnetic range.  


\subsec{Description of the Model}

The matter content of the models considered here
consists of two copies of
supersymmetric QCD, each with independent rank and
number of flavors, and a single gauge singlet chiral superfield
\eqn\fieldcontent{
\matrix{
\quad     & SU(N_c) & SU(N_c^{\prime})   &   \quad      \cr
          &         &             &  \quad       \cr
Q_i         & \fund   &    1       & i=1,\dots,N_f      \cr
\Qbar_i     & \overline{\fund} &  1  & \quad         \cr
P_{i^{\prime}}        &  1 &    \fund         &
    i^{\prime}=1,\dots,N_f^{\prime}      \cr
\Pbar_{i^{\prime}}      & 1 &  \overline{\fund} & \quad         \cr
\Phi      &    1    &    1        &   \quad      \cr
}}
The most general tree-level superpotential with only relevant or
marginal terms in four dimensions
for the matter content $\fieldcontent$ with $N_c$, $N_c'$ $\ge$ 4 
is
\eqn\Wtreegeneral{
W = (\lambda_{ij}  \Phi + \xi_{ij})   Q_i \Qbar_j +
  ( \lambda^{\prime}_{i^{\prime}j^{\prime}} \Phi
  + \xi^{\prime}_{i^{\prime}j^{\prime}})
    P_{i^{\prime}} \Pbar_{j^{\prime}} + w(\Phi) \, ,
}
where $w(\Phi)$ is a cubic polynomial in $\Phi$.  Remarkably, we shall find 
metastable vacua even in the simplest case of $w(\Phi)=0$, which we assume from now on.
The general situation is discussed in Section 5 (in $\felix$, 
the case $w(\Phi)=\kappa\Phi^3$ was used to stabilize $\Phi$ supersymmetrically).
 
At the classical level, the superpotential with $w(\Phi)=0$ has an
$U(1)_R \times U(1)_V \times U(1)_V^{\prime}$ global symmetry
under which the fields transform as
\eqn\treechargesgeneral{
\matrix{
\quad                & U(1)_R  & U(1)_V & U(1)_V^{\prime}       \cr
                     &         &        &       \cr
Q_i                  & +1       & +1      & 0     \cr
\Qbar_i              & +1       &  -1    & 0     \cr
P_{i^{\prime}}       & + 1      & 0      & +1    \cr
\Pbar_{i^{\prime}}   & +1       &  0     & -1 \cr
\Phi                    &    0    & 0      &  0  \cr
\Lambda^{3N_c - N_f} & 2N_c    &  0     &  0    \cr
\Lambda^{\prime 3N_c^{\prime} - N_f^{\prime}}
                   & 2N_c^{\prime}    &  0     &  0    \cr
}}
where
the normalizations of the $U(1)_V \times U(1)_V^{\prime}$ charges
are arbitrary.
In the quantum theory the $U(1)_R$ symmetry is anomalous
with respect to the  $SU(N_c)$ and $SU(N_c^{\prime})$ gauge dynamics.
The theta angles $\theta$ and $\theta^{\prime}$ transform
inhomogenously under $U(1)_R$, and the holomorphic dynamical scale,
\eqn\lamdef{
\left( \Lambda / \mu \right)^{3N_c - N_f} =
e^{-8 \pi^2 / g^2(\mu) + i \theta}\,,
}
and likewise for $\Lambda^{\prime 3N_c^{\prime} - N_f^{\prime}}$,
transform with charges given in~\treechargesgeneral.
The $U(1)_R$ symmetry is broken explictly by the anomalies to the
anomaly free discrete subgroups $Z_{2N_c} \subset U(1)_R$ and
$Z_{2N_c^{\prime}} \subset U(1)_R$, respectively.
The largest simultaneous subgroup of both
$Z_{2N_c}$ and $Z_{2N_c^{\prime}}$ which is left invariant
by the superpotential \Wtreegeneral~which couples the two
gauge sectors through $\Phi$ interactions is
$Z_{{\rm GCD}(2N_c , 2N_c^{\prime})} \subset U(1)_R$, where
${\rm GCD}(2N_c , 2N_c^{\prime})$ is the greatest common divisor
of $2N_c$ and $2N_c^{\prime}$.

In the $SU(N_f)_V \times SU(N_f^{\prime})_V$ global symmetry
limit the superpotential $\Wtreegeneral$ (with $w(\Phi)=0$) reduces to
\eqn\Wtreem{
W = (\lambda \Phi + \xi) {\rm tr}(   Q \Qbar) +
  (\lambda^{\prime} \Phi + \xi^{\prime})  {\rm tr}
    (P \Pbar )\,.
}
This superpotential has the same 
$U(1)_R \times U(1)_V \times U(1)_V^{\prime}$ global symmetry
as \Wtreegeneral, as well as a
$Z_2 \times Z_2$ conjugation symmetry under which
$Q_i \leftrightarrow \Qbar_i$ and
$P_i \leftrightarrow \Pbar_i$, respectively.
The form of the superpotential \Wtreem~may be enforced
for any $N_c$ and $N_c^{\prime}$ by weakly gauging
the $SU(N_f)_V \times SU(N_f^{\prime})_V$ symmetry.
One of the masses, $\xi$ or $\xi^{\prime}$, may always
be absorbed into a shift of $\Phi$.
For $\xi=\xi^{\prime}$ both masses may simultaneously
be absorbed into a shift of $\Phi$, and the tree level superpotential
in this case reduces to
\eqn\Wtree{
W = \lambda \Phi  ~{\rm tr}(   Q \Qbar) +
  \lambda^{\prime} \Phi~   {\rm tr}
    (P \Pbar )\,.
}
This form agrees with the naturalness requirement that there be no
relevant couplings.  $\Phi=0$ is an enhanced symmetry point for both sectors, where the 
respective quarks become massless. The case $\xi \neq \xi^{\prime}$ is analyzed in Section 5.

At the classical level this superpotential has an 
$U(1)_R \times U(1)_A \times U(1)_V \times U(1)_V^{\prime}$
global symmetry
\eqn\treecharges{
\matrix{
\quad              & U(1)_R & U(1)_A        &  U(1)_V &  U(1)_V^{\prime} \cr
                   &        &               &         &         \cr
Q_i                & +1    &  -{1 \over 2} &  +1     &  0      \cr
\Qbar_i            & +1    &  -{1 \over 2} &  -1     &  0        \cr
P_{i^{\prime}}     & +1    &  -{1 \over 2} &  0      &  +1  \cr
\Pbar_{i^{\prime}} & +1    &  -{1 \over 2} &  0      &  -1   \cr
\Phi                  & 0     &    +1         &  0      &   0 \cr
\Lambda^{3N_c - N_f} & 2N_c   &  -N_f   &  0     &  0    \cr
\Lambda^{\prime 3N_c^{\prime} - N_f^{\prime}}
                   & 2N_c^{\prime}  &  -N_f^{\prime}   &  0     &  0    \cr
       &          &       &               &     \cr
}}
where the normalizations of the $U(1)_A \times U(1)_V \times U(1)_V^{\prime}$
charges are arbitrary.
The $U(1)_R$ charges
are only defined up to an
addition of an arbitrary multiple of the $U(1)_A$ charges.
In the quantum theory both the $U(1)_R$ and
$U(1)_A$ symmetries are anomalous.
With the classical charge assignments \treecharges~the
$U(1)_R$ symmetry is broken explictly by the $SU(N_c)$
and $SU(N_c^{\prime})$ gauge dynamics to the anomaly
free discrete subgroup
$Z_{{\rm GCD}(2N_c , 2N_c^{\prime})} \subset U(1)_R$ as described above.
Likewise, the $U(1)_A$ symmetry is broken
explictly by $SU(N_c)$ and $SU(N_c^{\prime})$ gauge dynamics
to anomaly free discrete subgroups $Z_{N_f} \subset U(1)_A$ and
$Z_{N_f^{\prime}} \subset U(1)_A$, respectively.
The largest simultaneous subgroup of both
$Z_{N_f}$ and $Z_{N_f^{\prime}}$ which is left invariant
by the superpotential
\Wtree~is
$Z_{{\rm GCD}(N_f , N_f^{\prime})} \subset U(1)_A$.
The form of the potential ~\Wtree~may be enforced by gauging the
non-anomalous discrete $Z_{{\rm GCD}(N_f , N_f^{\prime})}$ symmetry if it
is non-trivial, along with weakly gauging the
$SU(N_f)_V \times SU(N_f^{\prime})_V$ symmetry.  This forbids the presence of 
a polynomial dependence $w(\Phi)$.

The marginal tree-level superpotential \Wtree~is,
up to irrelevant terms,
of rather generic form within many UV completions
of theories with moduli dependent masses.
It requires only that the masses of the flavors of both
gauge groups are moduli dependent functions,
and that all flavors become massless at a single point
in moduli space, here defined to be $\Phi=0$.
Importantly for the discussion of metastable dynamical supersymmetry
breaking below, the superpotential \Wtree~contains only marginal terms, so that
any relevant mass scales must arise from dimensional transmutation.
Generalizations to other
gauge groups and matter contents in vector-like
representations with the superpotential \Wtree~are straightforward.

The classical moduli space for the theory \fieldcontent~with
superpotential \Wtree~depends on the gauge group ranks
and number of flavors.
For $\lambda=\lambda^{\prime}=0$ the moduli space is
parameterized by $\Phi$, meson invariants
$M_{ij} = Q_i \Qbar_j$ and
$M^{\prime}_{i'j'} = P_{i'} \bar P_{j'}$
and for $N_f \geq N_c$ and/or $N_f^{\prime} \geq N_c^{\prime}$
baryon and anti-baryon invariants
$B_{i_1 i_2 \dots i_{N_c}} = Q_{[i_1} Q_{i_2} \cdots Q_{i_{N_c}]}$,
$\overline{B}_{i_1 i_2 \dots i_{N_c}} = \Qbar_{[i_1} \Qbar_{i_2}
  \cdots \Qbar_{i_{N_c}]}$, and/or
$B^{\prime}_{i_1 i_2 \dots i_{N_c'}} = P_{[i_1}
P_{i_2} \cdots P_{i_{N_c'}]}$,
$\overline{B}^{\prime}_{i_1 i_2 \dots i_{N_c'}} = \bar P_{[i_1}
\bar P_{i_2} \cdots \bar P_{i_{N_c'}]}$ respectively.
For $\lambda, \lambda^{\prime} \neq 0$ the superpotential \Wtree~lifts
all the moduli parameterized by the mesons.
The remaining moduli space
has a branch parameterized by $\Phi$.
For $\Phi \neq 0$ the flavors are massive and the baryon
and anti-baryon directions are lifted along this branch.
For $N_f \geq N_c$ and/or $N_f^{\prime} \geq N_c^{\prime}$
there is a second branch of the moduli space
parameterized by the baryons and anti-baryons
with $\Phi=0$.
The two branches touch at the point where all the moduli vanish.


\subsec{Supersymmetric Vacua}

The classical moduli space of vacua is lifted by nonperturbative
effects in the quantum theory.
Since the metastable supersymmetry breaking
vacua discussed below arise for
$\Phi \neq 0$, only this branch of the moduli space will be considered
in detail.
On this branch, holomorphy, symmetries, and limits fix the exact
superpotential written in terms of invariants, to be
\eqn\Wexactfull{ \eqalign{
W &=  ~\lambda \Phi ~{\rm Tr} M +
(N_c - N_f) \left[ { \Lambda^{3N_c-N_f} \over
   {\rm det}~M } \right]^{1/(N_c-N_f)}
  \cr  &
+ ~\lambda^{\prime} \Phi  ~{\rm Tr}  M^{\prime}  +
(N_c^{\prime} - N_f^{\prime})
  \left[ { \Lambda^{\prime 3N_c^{\prime}-N_f^{\prime}} \over
   {\rm det}~M^{\prime} }
   \right]^{1/(N_c^{\prime}-N_f^{\prime})}
   \cr
}}
For gauge sectors in the free magnetic range, the nonperturbative contribution
refers to the Seiberg dual.
Since the meson invariants are lifted on this branch, they may
be eliminated by equations of motion,
$\partial W / \partial M_{ij}=0$ and
$\partial W / \partial M^{\prime}_{i^{\prime}j^{\prime}}=0$,
to give the exact superpotential in terms of the classical
modulus $\Phi$
\eqn\Wexactgaugino{
W = N_c \left[ (\lambda \Phi)^{N_f}  \Lambda^{3N_c-N_f}
    \right]^{1/N_c}
+ N_c^{\prime} \left[ (\lambda^{\prime} \Phi)^{N_f^{\prime}}
   \Lambda^{\prime 3N_c^{\prime}-N_f^{\prime}}
    \right]^{1/N_c^{\prime}}\,.
}

The supersymmetric minima are given by stationary
points of the superpotential, $\partial W/ \partial \Phi =0$,
for which
\eqn\Sexact{
 N_f \left[ (\lambda \Phi)^{N_f}  \Lambda^{3N_c-N_f}
    \right]^{1/N_c}
+ N_f^{\prime} \left[ (\lambda^{\prime} \Phi)^{N_f^{\prime}}
   \Lambda^{\prime 3N_c^{\prime}-N_f^{\prime}}
    \right]^{1/N_c^{\prime}} =0\,.
}
Physically distinct supersymmetric vacua are distinguished by the expectation
value of the superpotential. 


\subsec{Parameter ranges for the gauge sectors}

Under mild assumptions we thus end up considering two SQCD sectors, characterized
by $(N_c, N_f, \Lambda)$ and $(N_c', N_f', \Lambda')$, respectively, and superpotential couplings
$\Wtree$. Different choices may be considered here; to restrict them, it is important
to note that calculable quantum corrections can be generated in two
different limits.

For $\lambda_i \Phi \gg \Lambda_i$, with $\Lambda_i = \Lambda$ or $\Lambda'$, the corresponding gauge
group is weakly coupled and hence generates small calculable corrections to the K\"ahler
potential. Integrating out the massive quarks, for energies below $\Phi$, leads to
gaugino condensation, which gives nonperturbative contributions as in \Wexactgaugino~.

On the other hand, for $\lambda_i \Phi \ll \Lambda_i$, the corresponding gauge sector becomes
strongly coupled. The calculable case corresponds to having the gauge theory in the free
magnetic range. For concreteness, we choose this sector to be $SU(N_c)$ (the unprimed sector), so
that $N_c+1 \le N_f < {3 \over 2} N_c$.

For the $(N_c', N_f', \Lambda')$ (primed) sector, 
the interesting case arises for $N_f'<N_c'$ and $\lambda' \Phi \gg \Lambda'$.
Although the classical superpotential pushes $\Phi$ to zero, the primed dynamics generate a 
nonperturbative term which makes the potential energy diverge as
$\Phi \to 0$, in agreement with the fact that $\Phi = 0$ corresponds to an enhanced symmetry point where $P$
and $\bar P$ become massless. Balancing the primed and unprimed contributions leads to a 
runaway direction in moduli space which will be lifted by one loop corrections.  
This stabilizes $\Phi$ at a nonzero value. 
Calculability demands working in the energy range $E \gg \Lambda'$
and $E \ll \Lambda$ so the dynamically generated scales must satisfy $\Lambda ' \ll \Lambda$.

The semiclassical limit corresponds to energies $E \gg \Lambda, \Lambda'$, where both sectors
are weakly coupled. Since $\Lambda' \ll \Lambda$, $SU(N_c)$ confines first when flowing to the IR.
For $\Lambda' \ll E \ll \Lambda$, the primed sector is weakly interacting while the unprimed sector 
has a dual weakly coupled description $\sem$ in terms of the magnetic
gauge group $SU(\tilde N_c)$ with $\tilde N_c=N_f-N_c$, $N_f^2$ singlets $M_{ij}$, and $N_f$
magnetic quarks $(q_i,\, \tilde q_j)$. In terms of this description, the full nonperturbative
superpotential reads
\eqn\Wclnp{ \eqalign{
W &= ~m\Phi~ {\rm tr}~M + h~{\rm tr}~ q M \tilde q +  \lambda'\Phi~ {\rm tr}~ P \bar P
+(N'_c - N'_f) \left( { \Lambda'^{3N'_c - N'_f}} \over {{\rm det}\, P \bar P} \right) ^{1/(N'_c - N'_f)}
  \cr &
+(N_f-N_c)\left({{\rm det} \,M \over \tilde \Lambda^{3N_c - 2N_f}}\right)^{1/(N_f-N_c)}.
  \cr
}}
Hereafter, $M_{ij} = Q_{i} \bar Q_{j}/\Lambda$, and  $m:=\lambda \Lambda$. 
The magnetic sector has a Landau pole at $\tilde{\Lambda}$ = $\Lambda$.  

In this description, the meson $M$ and the primed quarks $(P, \,\bar{P})$ become massless at $\Phi=0$.  
$M=0$ is also an enhanced symmetry point since here the magnetic quarks $(q, \, \tilde{q})$ become massless.


\newsec{Metastability near enhanced symmetry points}

In this section, metastable vacua near the origin of moduli space will be shown to exist for the theory
with superpotential $\Wclnp$. In Section 3.1,
we analyze the branches of the moduli space and determine where Coleman-Weinberg effects may lift the runaway. 
Next, in 3.2, we focus on the region containing metastable vacua.  In 3.3, we argue that other quantum 
corrections are under control and do not affect the stability of these vacua.  Finally, in Section 3.4 
the metastable vacua are shown to be parametrically long-lived.

\subsec{Exploring the moduli space}

Starting from the superpotential $\Wclnp$, the discussion is simplified by taking the limit 
$\tilde \Lambda \to \infty$, while keeping $m$ fixed.
The nonperturbative ${\rm det}\, M$ term is only relevant for generating supersymmetric vacua,
as discussed in $\Wexactgaugino$, and not important for the details of the metastable vacua that 
will arise near $M=0$.
Thus, for $M/\tilde \Lambda \rightarrow 0$ and $\Phi /\tilde \Lambda \rightarrow 0$, it
is enough to consider the superpotential
\eqn \Wfullmagn{
W = m\Phi~ {\rm tr}~M + h~{\rm tr}~q M \tilde q + \lambda'\Phi \,{\rm tr}\, P \bar P
+ (N'_c - N'_f) \left( { \Lambda'^{3N'_c - N'_f}} \over {{\rm det}\,P \bar P} \right) ^{1/(N'_c - N'_f)}\,.
}
In this limit all the fields are canonically normalized and the classical potential is
\eqn\Vcl{
V=V_D+V_D'+\sum_a |W_a|^2\,
}
where $W_a=\partial_a W$, and $a$ runs over all the fields. $V_D$ and $V_D'$ are the usual D-term contributions
from $SU(\tilde N_c)$ and $SU(N_c')$. Since both gauge sectors are weakly coupled, it is enough to consider the
F-terms on the D-flat moduli space, parametrized by the chiral ring. This restriction has no impact on
the analysis of the metastable vacua.

Let us study the regime $P \bar P \to \infty$.  Then nonperturbative effects from $SU(N_c')$ may
be neglected, and the classical superpotential
\eqn \Wclmagn{
W_{cl}=m\Phi~{\rm tr}~M + h~ {\rm tr}~q M \tilde q + \lambda'\Phi \,{\rm tr}~P\bar P
}
is recovered.
Setting
\eqn \Wx{
W_{M_{ij}}=m\Phi \delta_{ij}+h q_i \tilde q_j=0\,,
}
we obtain $\Phi =0$ and $h q \tilde q=0$.
This implies $ W_{{\rm tr}P\bar P} = W_q = 0$.  
The locus $ W_{\Phi} = 0$ then defines a classical moduli space of supersymmetric
vacua. 

Keeping $P\bar{P}$ large, but including the nonperturbative effects from $SU(N_c')$,  $W_{{\rm tr}P\bar P}=0$ sets 
$P\bar{P}\rightarrow\infty$ and $W_{\Phi}=0$ implies $M\rightarrow\infty$.
Therefore the model does not have a stable vacuum in the limit $\tilde{\Lambda}$ $\rightarrow$ $\infty$.
As discussed above, for $\tilde \Lambda$ finite and $M$ large enough, the nonperturbative ${\rm det}M$ term introduces
supersymmetric vacua as in \Wexactgaugino.

All the F-terms are small in the limit $M\to\infty$, $\Phi\to 0$, which thus corresponds to $M_F^2 \gg |F|$. The one-loop corrections
give logarithmic dependences on the fields $(\Phi, M)$ and these cannot stop the power-law runaway behavior.

Thus we are led to consider the region near the enhanced symmetry point $M=0$.  As we shall see below, 
this still has a runaway.  
Crucially, it turns out that one-loop corrections stop this runaway (this novel effect is characterized as  
a ``pseudo-runaway''). 
The reason for this is that the Coleman-Weinberg formula $\CW$
\eqn\Vcworiginal{
V_{CW}={1 \over {64 \pi^2}}\,{\rm Str} \,  M^4~ {\rm ln}~ M^2
}
will have polynomial (instead of logarithmic) dependence. This will be explained next.

\Ifig{\Fig\globalplot}{A plot showing the global shape of the potential.
$M$ has been expanded
around zero as in equation (3.8).  Note the runaway in the direction $X\rightarrow -\infty$ and
$\phi\rightarrow 0$.
The singularity at $\phi = 0$ and the ``drain'' $W_{\phi}=0$ are clearly
visible. Also visible is the Coleman-Weinberg channel near $X=0$ and $\phi$ large, discussed later.
This plot was generated with the help of $\Korneel$.}
{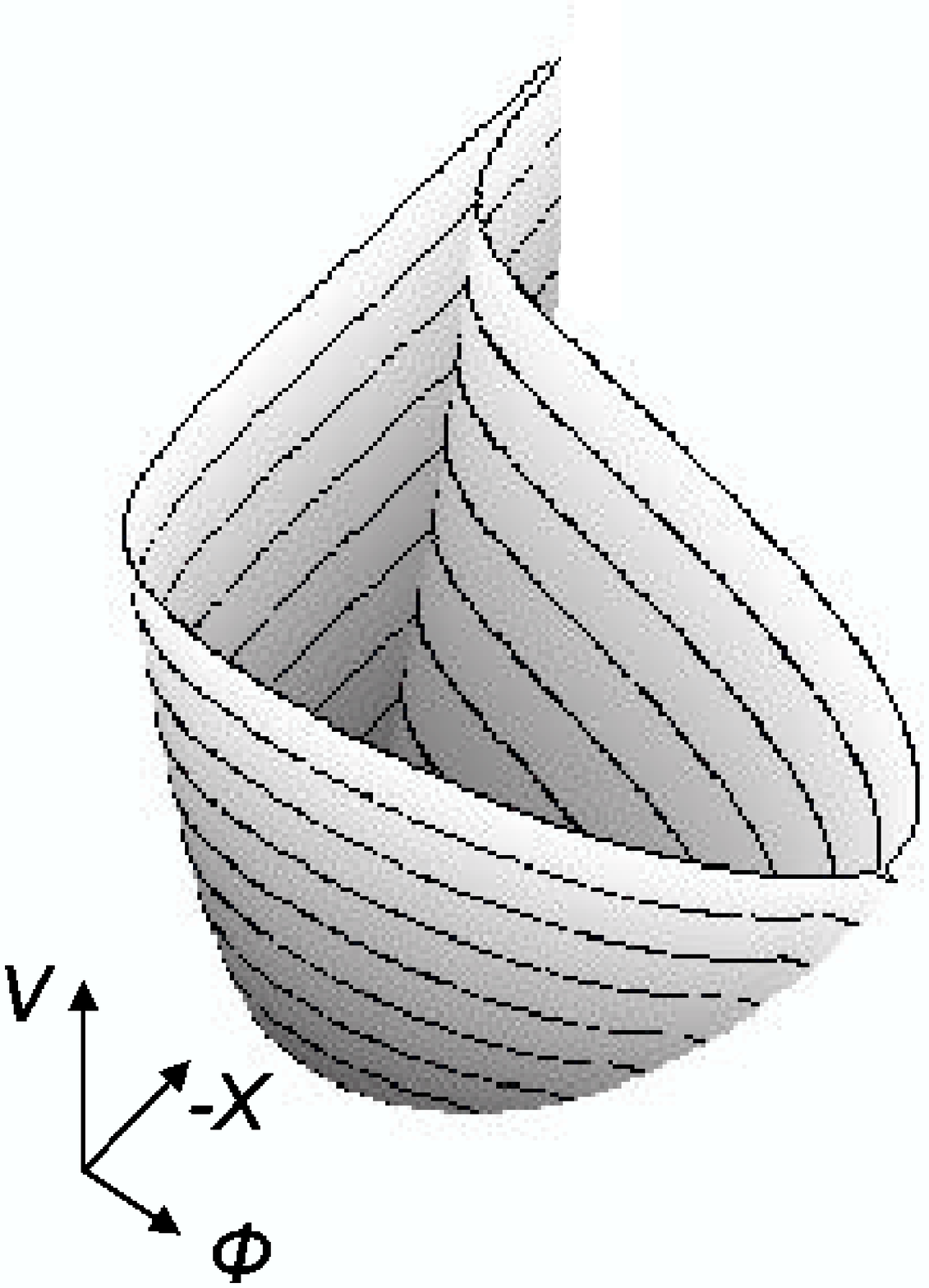}{3}
A global plot of the potential is provided in \globalplot, where $M$ has been expanded 
around zero as below in equation (3.8).
In the graphic, the `drain' towards the supersymmetric vacuum corresponds to the curve 
$W_\Phi=0$.


\subsec{Metastability Along the Pseudo-Runaway Direction}

In the region $\Phi \neq 0$, $(P, \bar P)$ may be integrated out by equations of motion provided that $\Lambda' \ll \lambda' \Phi$. This is a good 
description if we are not exactly at the origin but near it, as given by $\Phi / \tilde \Lambda \ll 1$. Taking, as before,
$\tilde{\Lambda} \to \infty$ and $m$ fixed, the superpotential reads
\eqn\Wdimple{
W=m\Phi\, {\rm tr}\,M+h \,{\rm tr}\,q M \tilde q+N_c' \big[\lambda'^{N_f'}
\,\Lambda'^{3N_c'-N_f'}\,\Phi^{N_f'}\big]^{1/N_c'}\,.
}
This description corresponds to
an O`Raifeartaigh-type model in terms of magnetic variables but with no
flat directions. 

Given that $\phi=\langle \Phi \rangle \neq 0$, we will expand around the point of maximal symmetry
\eqn\issvac{
q=\pmatrix{q_0&0}\;,\;\tilde q=\pmatrix{\tilde q_0 \cr 0}\;,\;M=\pmatrix{0&0\cr0& 0+X \cdot I_{N_c\times N_c}}\,.
}
Here $q_0$ and $\tilde q_0$ are $\tilde N_c \times \tilde N_c$ matrices satisfying
\eqn\qo{
h q_{0i} \tilde q_{0j}=-m\phi\,\delta_{ij}\;,\;i,j=\tilde N_c+1,\ldots N_f\,,
}
and the nonzero block matrix in $M$ has been taken to be proportional to the
identity; indeed, only ${\rm tr}\,M$ appears in the potential. This minimizes $W_M$ and
sets $W_q=W_{\tilde q}=0$. The spectrum of fluctuations around $\issvac$ is studied in detail in Section 4,
where it is shown that the lightest degrees of freedom correspond to $(\phi,  X)$ with
mass given by $m$. The effective potential derived from $\Wdimple$ is
\eqn\Vdimple{
V(\phi, X)=N_c m^2 |\phi|^2+\Bigg| m N_c X+N_f' \lambda'^{N_f'/N_c'}
\left({\Lambda'^{3N_c'-N_f'} \over \phi^{N_c'-N_f'}}\right)^{1/N_c'}\Bigg|^2+
V_{CW}(\phi, X)\,,
}
where the second term comes from $W_\phi$. 
The Coleman-Weinberg contribution will be discussed shortly.

As a starting point, set $ X=0$ and $V_{CW} \to 0$. Minimizing $V(\phi,  X=0)$ gives
\eqn\pdimple{
|\phi_0|^{(2N_c'-N_f')/N_c'}=\sqrt{{N_c'-N_f'} \over {N_c N_c'}}\,\,N_f' {\lambda'^{N_f'/N_c'} \over m} \Lambda'^{(3N_c'-N_f')/N_c'}\,,
}
and since $W_{\phi \phi} \sim m$, $V(\phi_0+\delta \phi, X=0)$ corresponds to a parabola of
curvature $m$. The nonperturbative term only affects $\phi_0$ but not the curvature $m$; this will be important in the discussion of
subsection 3.4. 

Next,
allowing $X$ to fluctuate (but still keeping $V_{CW} \to 0$), $V(\phi_0, X)$
gives a parabola centered at
\eqn\Xdrain{
X_{W_{\phi}=0} = - {\sqrt{N_c'\over{N_c(N_c'-N_f')}}}|\phi_0|
}
and curvature $m$. In other words, $X=0$ is on the side of a hill of curvature $m$ and height
$V(\phi_0, 0) \sim m^2 |\phi_0|^2$.

To create a minimum near $X=0$, $V_{CW}$ should contain a term $m_{CW}^2 |X|^2$,
with $m_{CW} \gg m$; this would overwhelm the classical curvature. As explained in Section 4, the massive degrees of freedom giving the dominant 
contribution
to $V_{CW}$ come from integrating out the massive fluctuations along $q_0$ and $\tilde q_0$. The result is
\eqn\Vcw{ V_{CW}=N_c bh^3 m |\phi| |X|^2+\ldots}
with $b=({\rm log}4-1)/8\pi^2 \tilde N_c$ $\iss$, and `$\ldots$' represent contributions that are unimportant for the present discussion. In this 
computation, $X$ and $\phi$ are taken
as background fields. It is crucial to notice that the quadratic $X$ dependence appears because
$X=0$ is an enhanced symmetry point.

In order to
be able to produce a local minimum, the marginal parameters $(\lambda ,\lambda')$ will have to be tuned to satisfy
\eqn\branch{ 
\epsilon \equiv {{m} \over m_{CW}} = {m\over bh^3 |\phi|} \ll 1 \,.
}
In this approximation, the value of $\phi$ at the minimum is still given by $\pdimple$; also, $X$
is stabilized at the nonzero value
\eqn\Xphi{
X_0=-e^{-i{{N_c'-N_f'}\over N_c'}\alpha_\phi}\,{N_f'\over bh^3} \lambda'^{N_f'/N_c'}\left({\Lambda'^{3N_c'-N_f'} \over
|\phi_0|^{2N_c'-N_f'}}\right)^{1/N_c'}.
}
The phases of $\phi$ and $X$ are thus related by
\eqn\phases{
\alpha_X+{{N_c'-N_f'} \over N_c'}\alpha_\phi=\pi\,.
}
Inserting $\pdimple$ into $\Xphi$ gives
\eqn\Xdimple{
|X_0|=\sqrt{N_c N_c' \over{N_c'-N_f'}}\,{m \over {bh^3}}\,.
}
\Ifig{\Fig\dimpleplot}{A plot showing the shape of the potential, including the
one-loop Coleman-Weinberg corrections, near the metastable minimum.
In the $\phi$-direction the potential is a parabola, whereas in the $X$-direction 
it is a side of a hill with a minimum created due to quantum corrections.  
This plot was generated with the help of $\Korneel$.}
{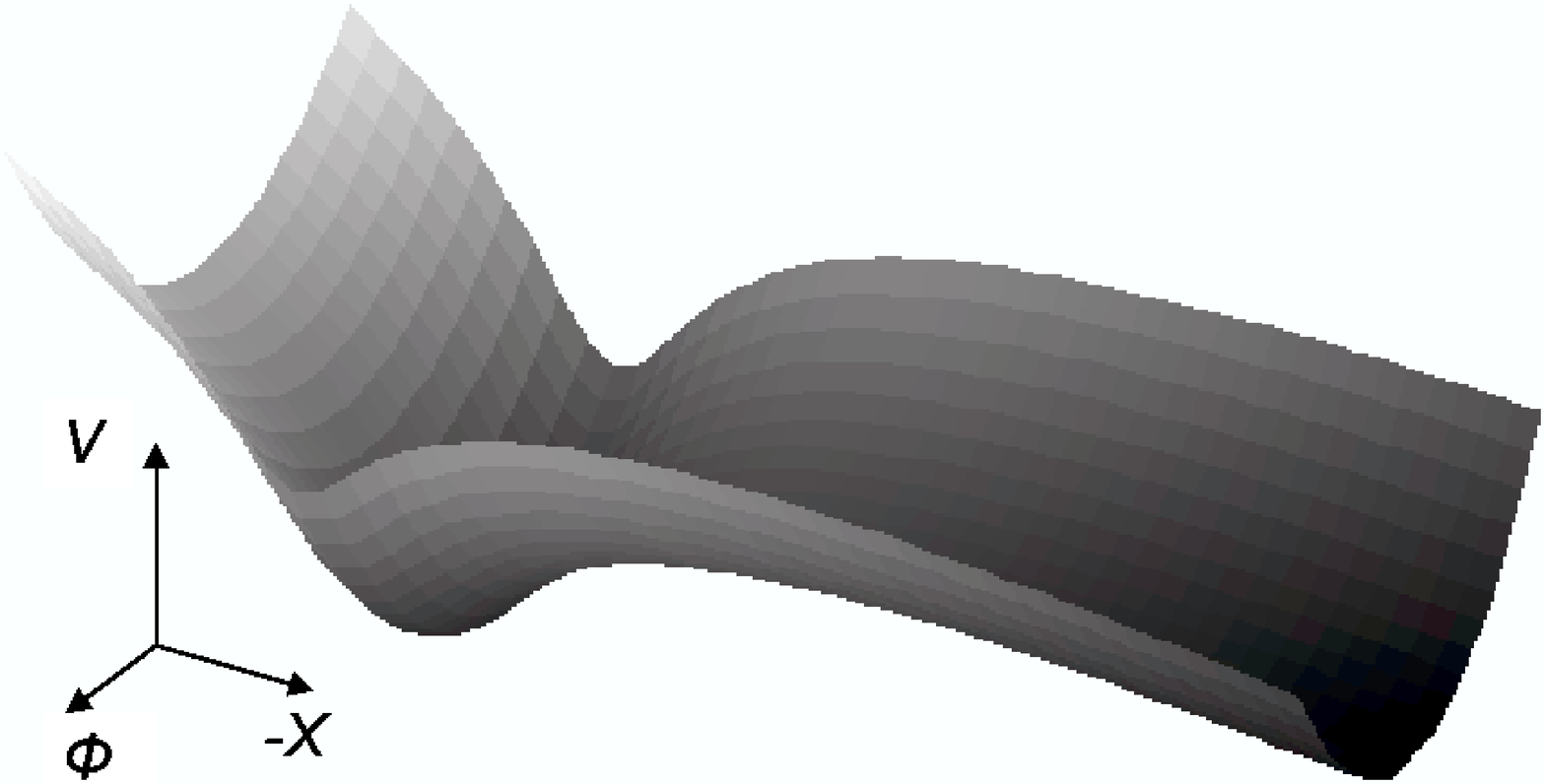}{5}

At the minimum, $\branch$ gives
\eqn\brancha{
(m/\Lambda')^{3N_c'-N_f'}\ll (bh^3)^{(2N_c'-N_f')/N_c'}\lambda'^{N_f'}
}
so the Yukawa coupling $\lambda$ in $m=\lambda \Lambda$ must be taken small for the analysis to 
be self-consistent.
The calculability condition $\Lambda' \ll \lambda' \Phi$ follows as a consequence of this.
At the minimum, $X_0 \ll \phi_0$.  The F-terms are given by 
\eqn\dimplevalues{
W_\phi \approx \sqrt{N_c N_c' \over{N_c'-N_f'}} m \phi_0 \sim W_X\;.
}
and from $\pdimple$ the scale of supersymmetry breaking is thus controlled 
by the dynamical scales of both gauge sectors.
In the next subsection, the vacuum will be shown to be long-lived if $\branch$ is satisfied.

Thus the model has a metastable vacuum near the origin, created by a combination of quantum corrections and
nonperturbative gauge effects.  The pseudo-runaway towards $X=X_{W_{\phi}=0}$ has been lifted by the 
Coleman-Weinberg contribution, as anticipated.  This is the origin of the $1/b$ dependence in $\Xdimple$.  The local 
minimum is depicted in 
\dimpleplot.


\subsec{Stability under other quantum corrections}
                                                                                                                                                             
The metastable vacuum appears from a competing effect between a runaway behavior
in the primed sector and one loop corrections for the meson field $X$. One is naturally led
to ask if, under these circumstances, other quantum effects are under control. These include higher loop
terms from the massive particles producing $V_{CW}$ as well as perturbative $g'$ corrections.

Let us first study higher loop contributions from the massive fields in $(q, \tilde q)$. They can correct the potential by additive 
terms of the form $X^n$, $n>2$; these are automatically subleading, because $|X_0|^2 \ll m |\phi_0|$. They can also produce
higher $\phi$ powers. However, such quantum corrections can only depend on the combination $m \phi$, and thus will be suppressed by powers of the
UV cutoff $\Lambda_0$. For instance, a quartic term would appear as $(m \phi)^4/\Lambda_0^4$. We conclude that all these effects are
subleading to $\Vcw$.
                                                                                                                                                             
Furthermore, since nonperturbative effects from $SU(N_c')$ were used, we should make sure that perturbative $g'$ effects are
not important. First note that the nonperturbative term
in $\Vdimple$ is of the same order as the classical height of the potential $m^2 |\phi|^2$ (see eq. $\dimplevalues$). It thus suffices to show
that $g'$ perturbative corrections to this height are subleading. A simple argument for this is as follows. Loops
generate typical quartic terms in the K\"ahler potential
\eqn\quartic{
\delta K= {{\alpha} \over {\Lambda_0^2}}(\Phi^* \Phi)^2
}
which change the scalar potential by 
\eqn\changequartic{
\Big[{{\alpha} \over {\Lambda_0^2}}\,|\phi|^2\Big]\,(m^2 |\phi|^2)\,.
}
The prefactor is parametrically small, making these contributions negligible.


\subsec{Tunneling Out of the Metastable Vacuum}

This section will show that the metastable non-supersymmetric vacuum
can be made parametrically long-lived by taking the parameter
$\epsilon \equiv {m \over {b h^3 |\phi_0|}}$ sufficiently small.
The lifetime of the metastable vacuum may be estimated using
semiclassical techniques and is proportional to the exponential
of the bounce action, $e^B$ $\coleman$.

First, the direction of tunneling in field space needs to be determined.
Recall that the metastable vacuum in the ($|\phi|$, $X$)
space lies at
\eqn\dimpleposition{
|\phi_0|^{(2N_c'-N_f')/N_c'}=\sqrt{{N_c'-N_f'} \over {N_c
N_c'}}\,\,N_f' {\lambda'^{N_f'/N_c'} \over m}
\Lambda'^{(3N_c'-N_f')/N_c'},\;\;\;\;\;\; X_0=-{\sqrt{N_c
N_c' \over{N_c'-N_f'}}}\,{m \over {bh^3}}\,. }
(The phase of $\phi$, not of qualitative importance for the present
discussion, has been chosen to be zero.  This fixes $X$ to be real - see equation $\phases$.)
For fixed $X$ the potential has a minimum at $|\phi| = |\phi_0|$;
while quantum corrections may change this value by an order
one number, corrections to the curvature of the potential
in the $|\phi|$ direction are negligible.
This curvature is positive, and thus the potential increases
as $|\phi|$ moves away from $|\phi_0|$.
The field therefore does not tunnel in the $|\phi|$ direction (see \dimpleplot).
Along the $X$ direction, however, the potential without quantum
corrections near the enhanced symmetry point is like the side of a hill.
For fixed $|\phi|=|\phi_0|$, the potential decreases in the negative
$X$ direction, and the classical curvature at $X=0$ is $m$.

Quantum corrections are qualitatively important when $|X|$
is sufficiently small.
For $|X|^2 \ll |W_X|$, their size grows quadratically as a
function of $X$ and they are sufficient to change the slope of the
classical potential enough to introduce a minimum.
For $|X|^2 \simeq |W_X|$, the growth of the quantum corrections
is only logarithmic, and the slope of the classical potential
again starts to dominate.
Hence, the total potential has a peak that parametrically may
be estimated to lie near
\eqn\Xpeak{
X_{{\rm peak}} \simeq -\sqrt{|W_X|} = -\sqrt{N_c m |\phi_0|}.
}
For $X > X_{\rm peak}$, the potential decreases as $X$
becomes more negative until $X$ reaches the `drain' $W_{\phi}=0$,
\eqn\Xdrain{
X_{W_{\phi}=0} = - {\sqrt{N_c'\over{N_c(N_c'-N_f')}}}|\phi_0|.
}
The direction in field space to tunnel out of the false vacuum
is towards negative $X$ with fixed $|\phi|=|\phi_0|$.
It thus suffices to consider the tunneling in the one-dimensional
potential, $V(X)$ $\equiv$ $V(|\phi_0|,X)$.
Note that parametrically $|X_0|$ $\ll$ $|X_{{\rm peak}}|$ $\ll$
$|X_{W_{\phi}=0}|$ as $\epsilon \rightarrow 0$.

For negative $X$, using equations $\Vdimple$ and $\dimpleposition$,
the one-dimensional potential may be written as
\eqn\Vrewritten{
V(X) = \Bigg({2N_c'-N_f' \over N_c'-N_f'}\Bigg)\,N_c\,m^2\,|\phi_0|^2
    + N_c^2\,bh^3\,m^2\,|\phi_0|^2\,f\bigg({-|X|\over bh^3|\phi_0|}\bigg).
}
In the region $X \ll X_{\rm peak}$, the function $f(x)$
is dominated by quantum corrections and may be approximated by
\eqn\fxsmallX{
f(x) \simeq {bh^3\over N_c\,\epsilon}\,x^2\,,
}
where a constant piece coming from the quantum corrections, again not important for 
the calculation of the bounce action, has been neglected.  
On the other hand, in the region $X_{\rm peak} \ll X \ll 
X_{W_{\phi}=0}$,
the constant slope of the classical potential dominates.
The potential in
this region may be approximated by the classical
potential plus a constant contribution from the
quantum corrections whose size is roughly given by the height of the potential barrier.
The height of the potential barrier is, from 
$\fxsmallX$, of order $f(X_{\rm peak}/bh^3|\phi_0|) = 1$, and it is thus
loop-suppressed compared to the overall magnitude of the potential near the metastable minimum.
The potential in this region will be parametrized by a straight line
\eqn\fxlargeX{
f(x) \simeq 1 - 2\sqrt{N_c'\over N_c(N_c'-N_f')}\,(x-x_{\rm peak}).
}

In order to estimate the bounce action it is not appropriate
to use the thin-wall approximation
$\coleman$.
Instead, the potential may be modeled as a triangular barrier $\duncan$.
Using the results of \duncan, the value to which the field tunnels to is
\eqn\tunnelvalue{
\tilde{X} 
        \sim 
            - \, b\,h^3|\phi_0|.
}
Note that parametrically $|X_0|$ $\ll$ $|X_{{\rm peak}}|$ $\ll$ $|\tilde{X}|$ as $\epsilon \rightarrow 0$,
and that $|\tilde{X}|$ is loop-suppressed compared to
$|X_{W_{\phi}=0}|$.
The bounce action scales as
\eqn\bounce{
B \sim 
    {{\tilde{X}^4}\over{V(X_{peak})-V(X_{0})}}
  \sim b\, h^3 \, {1\over \epsilon^2}.
}
Therefore $B\rightarrow\infty$ as $\epsilon\rightarrow 0$, and the
metastable vacuum is parametrically long-lived.

\Ifig{\Fig\OneDplot}{
A plot of the classical potential (dashed line) and the total potential
including one-loop corrections (solid line) for fixed $|\phi|=|\phi_0|$,
where $|\phi_0|$ is the position of the metastable minimum
in the $\phi$-direction, defined in $\dimpleposition$.
In the figure, $N_f=3$, $N_c=2$, $N_f'=1$ and $N_c'=2$.
The values were scaled so that the position of the ``drain'',
$W_{\phi}=0$, equals 1 on both axes.
In these units, the position of the metastable minimum is on
the order of $10^{-4}$.
This plot was generated with the help of $\Korneel$.}
{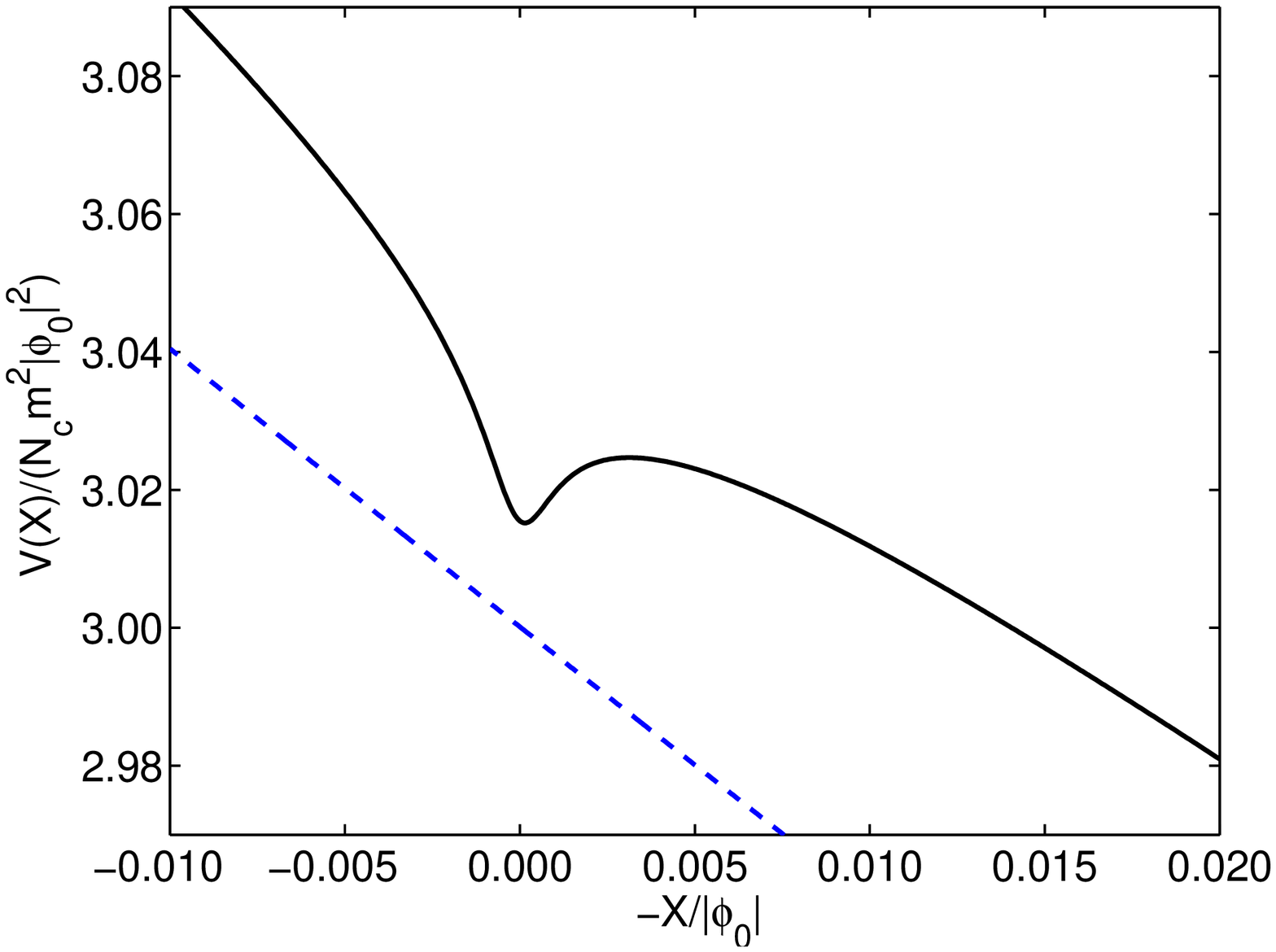}{5}
The total potential $V(X)$, including the full one-loop Coleman-Weinberg potential
computed numerically with the help of $\Korneel$, is shown in \OneDplot.  
The program of $\Korneel$
also allowed us to check numerically the previous tunneling properties.


\newsec{Particle Spectrum and R-symmetry}

In this section, we discuss in more detail the particle spectrum of the model and comment on the 
R-symmetry properties.

The fluctuations of the fields around the metastable minimum may 
be parametrized following ISS,
\eqn\particlespW{
\phi=\phi_0+\delta \phi\;,\;M=\pmatrix{Y_{\tilde N_c \times \tilde N_c}&Z^T_{\tilde N_c \times (N_f-\tilde N_c)} \cr \tilde Z_{(N_f-\tilde N_c)\times \tilde N_c} &
X_0 + X_{(N_f-\tilde N_c) \times (N_f-\tilde N_c)}}
}
\eqn\particlesq{
q=\pmatrix{q_{0} + \chi_{\tilde N_c \times \tilde N_c}\cr \rho_{(N_f-\tilde N_c)\times \tilde N_c}}\;,\;\tilde q=\pmatrix{\tilde q_{0} + \tilde \chi_{\tilde N_c \times \tilde N_c}\cr \tilde \rho_{(N_f-\tilde N_c)\times \tilde N_c}}\,,
}
where $q_{0} \tilde q_{0}:=-m \phi_0/h$. All fields are complex; $\phi_0$ and $X_0$ are the values at the metastable minimum.

The relevant mass scales are
\eqn\scalemass{
M^2=0,\,m^2,\,m_{CW}^2=bh^3 m |\phi_0|,\,hm|\phi_0|\,.
}
The particles may
be divided into three `sectors' with small mixing amongst themselves. Up to quadratic order, the superpotential is
\eqn\Wfluc{
\matrix{W&=&W_{\phi \phi} ~\delta \phi ~\delta \phi~+~mN_c ~ \delta \phi ~(X_0+ X)~+~m \delta \phi ~\sum_{\alpha=1}^{\tilde N_c} \,Y_{\alpha \alpha}+ \;\;\;\;\;\;\;
\;\;\;\;\;\;\; \cr
& +&mN_c ~\phi_0 ~(X_0+ X)~+~h\sum_{f=1}^{N_c}\,[~q_0~ (\tilde \rho Z^T)_{ff}+\tilde q_0 ~(\rho \tilde Z^T)_{ff}+ X_0 ~(\rho \tilde
\rho^T)_{ff}~]\cr
&+&h~ \sum_{\alpha=1}^{\tilde N_c}\, [q_0~(\tilde \chi Y)_{\alpha \alpha}+\tilde q_0(\chi Y)_{\alpha \alpha}~]\,.\;\;\;\;\;\;\;
\;\;\;\;\;\;\;\;\;\;\;\;\;\;
\;\;\;\;\;\;\;\;\;\;\;\;\;\;
\;\;\;\;\;\;\;\;\;\;\;\;\;\;
\;\;\;\;\;\;\;
}
}
The first line is related to the new dynamical field $\delta \phi$; 
unlike ISS, now $X$ is not a pseudo-flat direction. The second and third lines are as in ISS.

\Ifig{\Fig\massesTable}{
Table showing the classical mass spectrum, grouped in sectors of ${\rm Str} \,m^2=0$ for $N_f=N_c+1$. 
The $O(m^2)$ fields in $(\phi,\, {\rm tr}\,X)$ 
are not
degenerate. Although supersymmetry
is spontaneoulsy broken, there is no goldstino at the classical level.}
{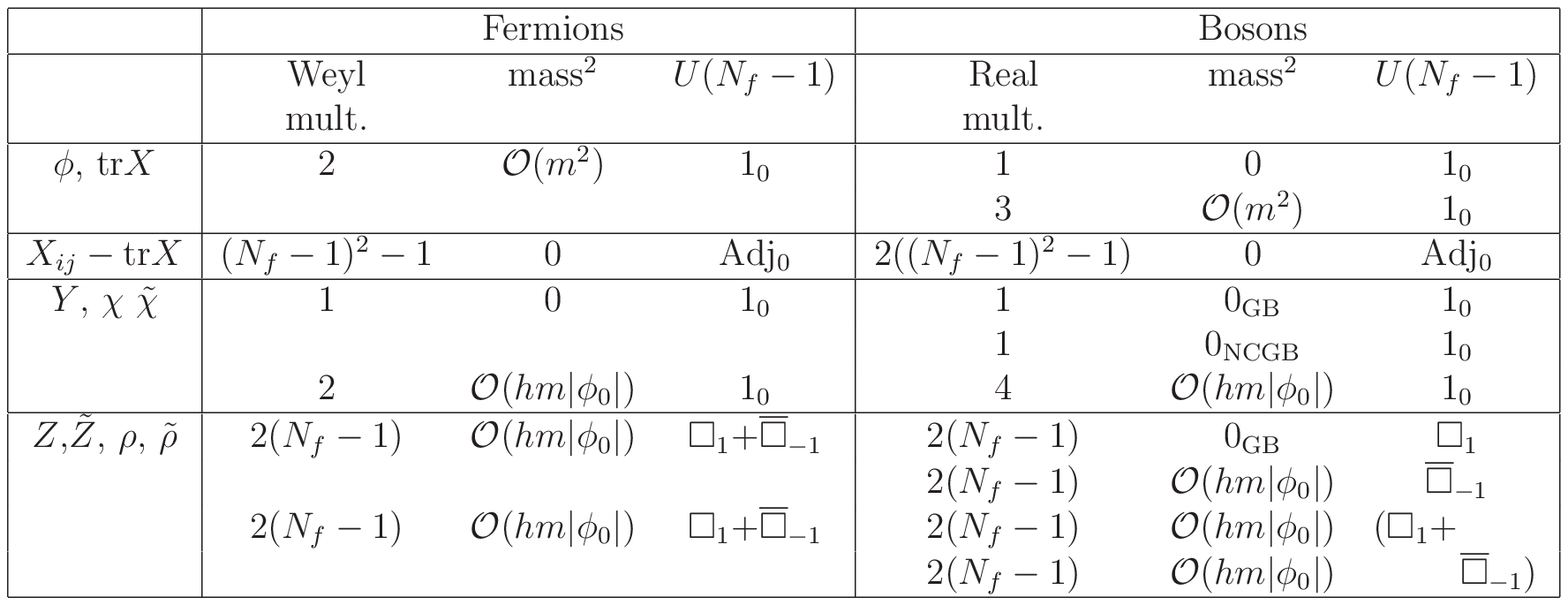}{6}

\Ifig{\Fig\massesTableCW}{
Table showing the mass spectrum, including one-loop corrections corrections, grouped in sectors of ${\rm Str} \, m^2=0$
for $N_f=N_c+1$. Notice the appearance
of the goldstino in the $(\phi, {\rm tr}\,X)$ sector. The $O(m^2)$ fields in $(\phi,\, {\rm tr}\,X)$ are not
degenerate; here $m_{CW}^2=bh^3 m|\phi_0|$.
}
{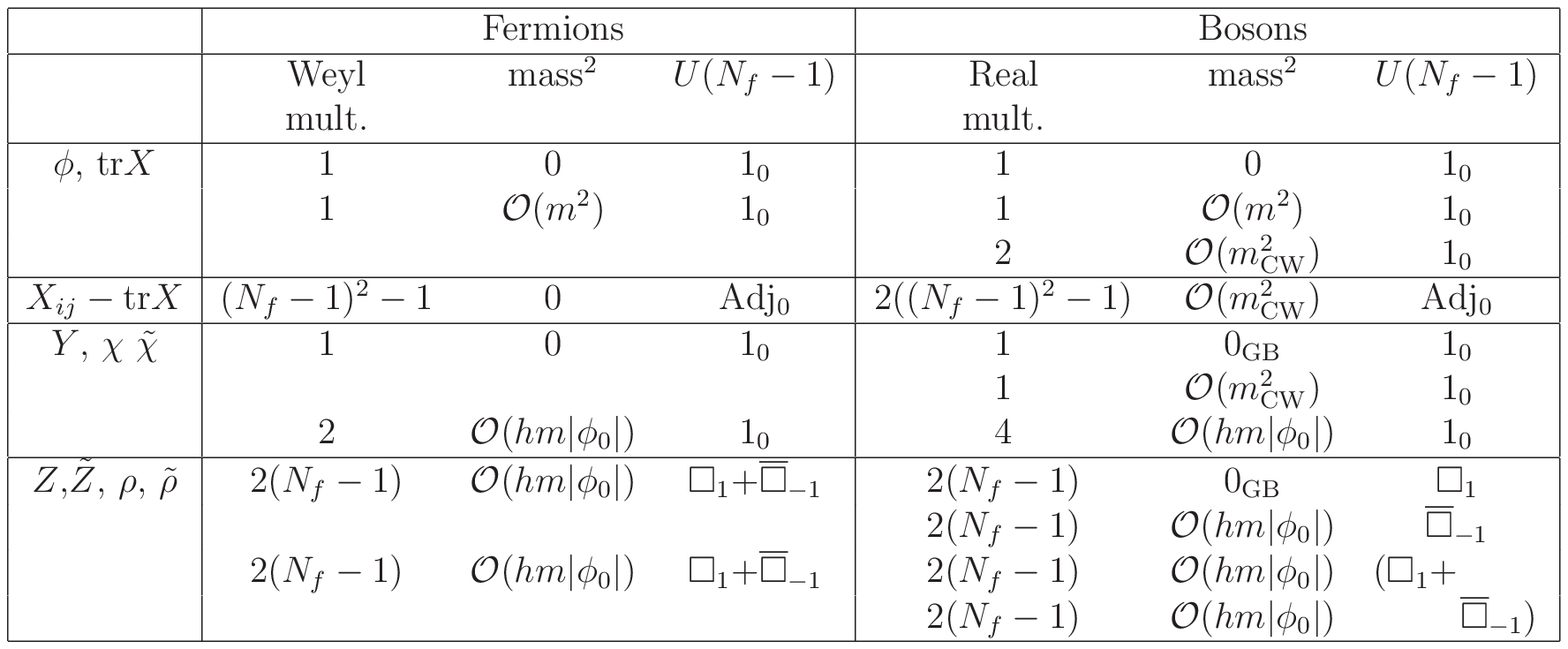}{6}

Consider the case $N_f=N_c+1$; the spectrum of classical masses is shown in \massesTable, and the spectrum of the masses including one-loop CW
corrections is shown in \massesTableCW. The fields are grouped in sectors of ${\rm STr} M^2=0$.

The fields $(Y, \chi, \tilde \chi)$ form three chiral superfields, with supersymmetric masses, and hence do not
contribute when integrated out at one loop. The Coleman-Weinberg potential is generated by the fields $(Z, \tilde Z, \rho, \tilde \rho)$, which
are the heaviest in the spectrum. Including such quantum corrections, ${\rm tr}\,X$ acquires a mass $m_{CW}^2$, while the mass of $\phi$
is not modified. Interestingly, at the classical level there is no massless goldstino, since the expansion is not around a critical point of
the classical potential. Including quantum corrections, one of the massive fermions in the $(\phi, {\rm Tr}\,X)$-sector becomes massless,
as may be seen in \massesTableCW. A similar situation, in the opposite limit of small supersymmetry breaking, has been discussed recently in~\review.

\Ifig{\Fig\massesTableN}{
Table showing the classical mass spectrum, grouped in sectors of ${\rm Str} \, m^2=0$, for $N_f>N_c+1$. After gauging $SU(\tilde N_c)$, the
traceless goldstone bosons from $(\chi, \bar \chi)$  are eaten, giving a mass $m_W^2=g^2 m|\phi_0|/h$ to the gauge bosons. Further, from $V_D=0$, the
noncompact goldstones also acquire a mass $m_W^2$. Including CW corrections, ${\rm tr}\,X$ acquires mass $m_{CW}^2$ and one of the fermions becomes massless.
}
{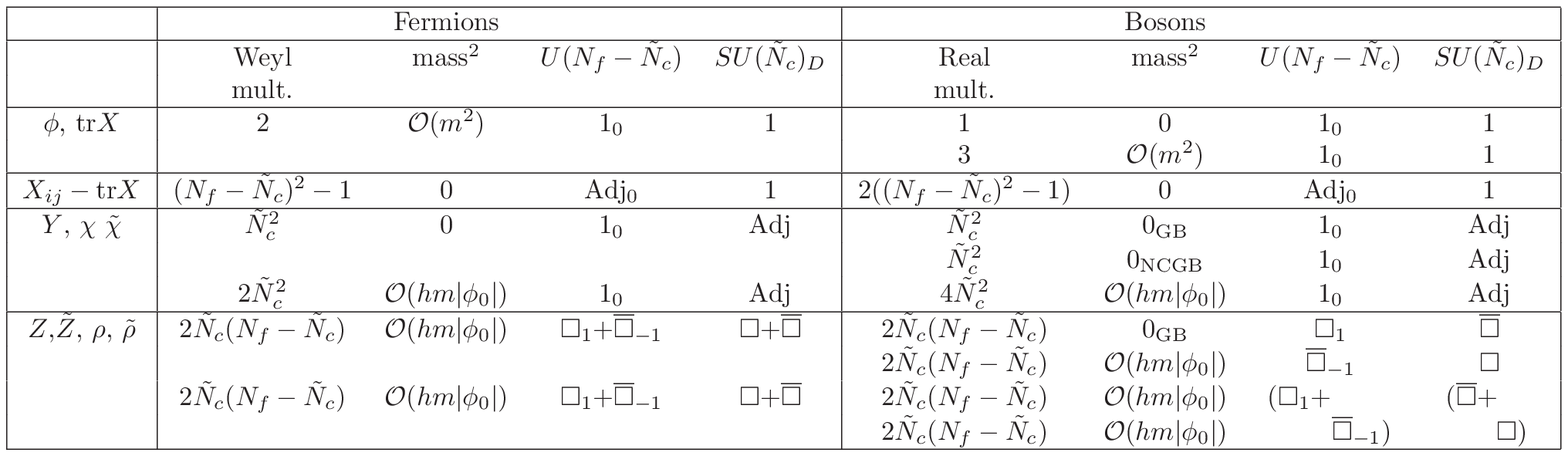}{7}

The case $\tilde N_c=N_f-N_c > 1$ can be similarly analyzed, and is shown in \massesTableN.

The Standard Model gauge group can be embedded inside the 
global symmetry group of this model. In this way, renormalizable models of direct gauge mediated supersymmetry breaking may be constructed.

\subsec{Breaking the R-symmetry}

To have gaugino masses, any R-symmetry must be
broken, explicitly and/or spontaneously $\iss$, $\review$. The low energy superpotential $\Wdimple$ has the following $U(1)_R$ symmetry:
\eqn\Rdimple{
R_{\phi}=2 {N_c' \over N_f'}\;,\;R_X=2 {{N_f'-N_c'} \over N_f'}\;,\;R_q=R_{\tilde q}={N_c' \over N_f'}\,.
}
Since the VEV's of these fields are nonzero in the metastable vacuum, the R-symmetry is spontaneously broken, and there is
an R-axion $a$. In terms of the phase of the $i$-th field, the axion is
\eqn\axion{
\phi_i={1 \over \sqrt 2}\,{f_R \over R_i}\,e^{i R_i(a/f_R)}\,,
}
where the decay constant $f_R$ is defined as
\eqn\fr{
f_R= \Big[\sum_i \big( {\sqrt 2} R_i |\langle \phi_i \rangle|\big)^2 \Big]^{1/2}
}
and $R_i$ is the R-charge of $\phi_i$. In $\shih$ it was pointed out that if R-symmetry is broken 
spontaneously in an O' Raifeartaigh model, then
the theory should contain a field with R-charge different than 0 or 2. 
This is also the case in the present situation, although our model does not contain
the linear O' Raifeartaigh term.

For finite $\tilde \Lambda$, the ${\rm det}~X$ contributions need to be taken into account, and the $U(1)_R$ symmetry
becomes anomalous. Adding this term induces a tadpole for $Y$, which now acquires an expectation value of order
$$
Y \sim \bigg[{X_0 \over \tilde \Lambda}\bigg]^{{3N_c-2N_f}\over{N_f-N_c}}\,X_0 \ll X_0\,.
$$
Then the mass of the R-axion follows from
$$
|W_X|^2 \sim \Bigg|m \phi + c X_0^2 \,\bigg[{X_0 \over \tilde \Lambda}\bigg]^{2\,{{3N_c-2N_f}\over{N_f-N_c}}} \Bigg|^2\,.
$$
Deriving twice the cross-term, which is proportional to ${\rm cos}(a/f)$, yields the axion mass
\eqn\axionm{
m_a^2 \sim m^2\,\Bigg(\bigg[{\lambda \over {bh^3}}\bigg]^{2\,{{3N_c-2N_f}\over{N_f-N_c}}}\,{\epsilon \over{bh^3}} \Bigg)\ll m^2\,,
}
where $\lambda$ is the Yukawa coupling appearing in $m=\lambda \Lambda$.
Thus, R-symmetry is both spontaneously and explicitly broken.


\newsec{Meta-Stability Near Generic Points of Enhanced Symmetry}

In this section, the existence and genericity of metastable vacua near enhanced symmetry points is explored. Statistical analyses
of the supersymmetry breaking scale up to date have not taken into account loop quantum effects $\mike$, as these corrections are
hard to evaluate on an ensemble of field theories. However, metastable
vacua introduced by the Coleman-Weinberg potential, with all the relevant parameters generated dynamically, may change such results.
Before considering the general case, let us analyze $\Wtreem$.


\subsec{Non-coincident enhanced symmetry points}

Consider two gauge sectors as in $\Wtreem$, with enhanced symmetry points at $\Phi=0$ and $\Phi=\xi$, respectively.
The free magnetic sector is taken to be massless at $\Phi=0$; integrating over the other primed sector gives
\eqn\Wdimplex{
W=m\Phi\, {\rm tr}~M+h\, {\rm tr}\,q M \tilde q+N_c' \big[\lambda'^{N_f'}
\,\Lambda'^{3N_c'-N_f'}\,(\Phi+\xi)^{N_f'}\big]^{1/N_c'}\,.
}
Since metastable vacua were shown to exist for $\xi=0$, here the discussion is restricted to the limit
of $\xi$ much bigger than all the energy scales in the problem. This is consistent with the fact that naturalness
demands any relevant coupling to be of order the UV cutoff.

Introducing the notation
$$
\alpha=N_f'/N_c'\;,\;K=N_c' \lambda'^{N_f'/N_c'}\,\Lambda'^{(3N_c'-N_f')/N_c'}\,,
$$
the equations of motion for $\phi$ and $X$ give
\eqn\exphi{
N_c m^2 \phi = \alpha^2 (1-\alpha) {K^2 \over {\xi^{3-2\alpha}}}\,.
}
\eqn\exX{
|X|={N_c \over{\alpha (1-\alpha)}}\,{{m^2 \xi^{2-\alpha}} \over K}\,.
}

Without fine-tuning $m$ or $K$, $X$ tends to be driven away from the origin as $\xi$ increases.  
The fine-tuning may be seen, for instance,
from the requirement $m_{CW} \gg m$, which implies
\eqn\extune{
m^3 \ll bh^3 {K^2 \over \xi^{3-2\alpha}}\,.
}
Although this resembles the calculability condition $\brancha$, now there are powers of the large
scale $\xi$ in the denominator.
For $\xi$ of order the UV cutoff, this represents a big fine-tuning, either on the coefficient $K$ or on the small
mass parameter $m$. 

The conclusion is that, while metastable vacua can occur for far away enhanced symmetry
points, this situation is not generic and requires fine-tuning. This is to be expected, once relevant parameters are allowed
to appear in the superpotential.


\subsec{General Analysis}

A generic structure in the landscape of effective field theories corresponds to a gauge theory with vector-like matter
and mass given by a singlet, whose dynamics is related to another sector. The superpotential may be written as
\eqn\genw{
W=f(\Phi)+\lambda\,\Phi {\rm tr}(Q \bar Q)\,.
}
Here, $(Q, \bar Q)$ are $N_f$ quarks in $SU(N_c)$ SQCD; $f(\Phi)$ may be generated, for instance, from a flux superpotential, by 
nonrenormalizable interactions $\retro$, or,
as in the case studied in this work, by another gauge sector. Next, it is required that the SQCD sector be in the free magnetic range;
this is still a generic situation. The dual magnetic description is weakly coupled near the enhanced symmetry point $\Phi=0$, where the superpotential reads
\eqn\genw{
W=f(\Phi)+m \Phi\, {\rm tr}~M + h\, {\rm tr}~q M \tilde q\,.
}

The question that will be addressed here is: what restrictions need to be imposed on $f(\Phi)$, so that the one loop potential
$V_{CW}$ can create a metastable vacuum near $M=0$?  Since we are interested in the novel effect of pseudo-runaway 
directions we will demand $f'(\Phi)\neq 0$.  The case $f'(\Phi)=0$ is standard in such analyses, see e.g. $\felix$.

As discussed in Section 3, this is possible only if
\eqn\restrm{
m_{CW}^2:=N_c bh^3 m |\phi| \gg m^2\,
}
where $\phi$ denotes the expectation value of $\Phi$ at the metastable vacuum. Further, one needs to impose that
\eqn\restrx{
h^2 |X|^2 \ll m |\phi|
}
in order for the Taylor expansion of $V_{CW}$ around $X=0$ to converge. Evaluating the potential as in \Vdimple,
\eqn\vgen{
V=N_c m^2 |\phi|^2+ \big|f'(\phi)+m N_c\, X \big| ^2+m_{CW}^2 |X|^2\,.
}
The rank condition, an essential ingredient in the discussion, just follows from having SQCD in the free magnetic range.
This fixes the first term, which comes from $W_M$, and the block structure of the matrix $M$; $X$ was defined in $\issvac$.

Extremizing $V(\phi, X=0)$ leads to
\eqn\phiextr{
N_c m^2 \phi=-f'(\phi)\, f''(\phi)^*\,.
}
On the other hand, minimization with respect to $X$ in the approximation $m_{CW}^2 \gg m^2$, gives the metastable vacuum
\eqn\Xf{
m_{CW}^2 \, X=-N_c m f'(\phi)\,.
}
Notice that $m_{CW}^2 \gg m^2$ makes this value parametrically smaller than the position of the `drain' $f'(\phi)+m N_c\, X=0$.
This ensures the stability of the nonsupersymmetric vacuum. Replacing $\phiextr$ in $\Xf$ (with $m_{CW}^2=N_c bh^3 |\phi|$) yields
\eqn\Xextr{
|X|={{N_c m^2} \over {bh^3}}\, {1 \over {|f''(\phi)|}}\,.
}

It is possible to combine the conditions $\restrm$ and $\restrx$
with the values at the metastable vacuum $\phiextr$, $\Xextr$, to derive constraints on $f(\phi)$: $\restrm$ now reads
\eqn\restrma{
{|f'(\phi)f''(\phi)| \over m^3} \gg {1 \over {bh^3}}\,,
}
while \restrx~gives
\eqn\restrxa{
h^2 |f'(\phi)|^2 \ll m (bh^3)^2 |\phi|^3\,.
}
Summarizing, the necessary conditions for metastable vacua near $X=0$ to exist are $\restrma$ and $\restrxa$. As illustrated in the
previous subsection, they require fine-tuning the coefficients of $f(\phi)$, except in the case of coincident enhanced symmetry points, where
there are no relevant scales.


\newsec{Conclusions}

In this paper we constructed a model with long-lived metastable vacua in which
all the relevant parameters, including the supersymmetry breaking scale, are generated dynamically
by dimensional transmutation.  The model consists of two $N=1$ supersymmetric QCD sectors with
flavors whose respective masses are controlled by the same singlet field.
One of the gauge sectors is in the free magnetic range while the other is in the electric range.
The metastable vacua are produced near a point of enhanced symmetry by a combination
of nonperturbative gauge effects and, crucially, perturbative effects coming from the one-loop
Coleman-Weinberg potential.

The model has the following desirable features: an explicitly and spontaneously broken 
$R$-symmetry, a singlet, a large global symmetry, naturalness and renormalizability.  

There are two points that have to be stressed. First, a salient feature of the model is the
existence of pseudo-runaway directions. They correspond to a runaway behavior that is lifted
by one loop quantum corrections. This has not been observed before, the closest analog corresponding
for example to the pseudo-moduli of $\iss$. It is quite plausible that this phenomenon appears in other models as well.
The criterion is that the height of the potential has to be parametrically larger than the curvature, as
quantified in Section 3. The strength of the quadratic Coleman-Weinberg corrections is set by this height,
thus introducing a local minimum of high curvature in the (otherwise) runaway potential.

In dynamical supersymmetry breaking models $\WittenDSB$, $\ADS$, nonsupersymmetric vacua generally arise due to competing effects between a 
nonperturbative runaway and a classical term in the superpotential, as in the (3,2) model $\three$. Our analysis shows that
it is possible to stabilize such runaways even without tree-level terms, provided that one is close to certain enhanced 
symmetry points.

The second feature worth emphasizing is the connection between enhanced symmetry points in gauge theory moduli spaces
and metastable dynamical supersymmetry breaking. There are reasons to believe that such vacua are generic. 
At the field theory level this is associated to the fact that a nonzero Witten index $\Wittenindex$ may still allow an approximate R-symmetry $\issa$.
While dynamical ISS models are not hard to construct, in general these mechanisms involve discrete R-symmetries $\retro$.
This is very suppressed in the landscape of string vacua, correponding to a high codimension locus in the flux lattice $\dinea$.
On the other hand, the construction presented here does not suffer from the previous difficulty. Therefore, it would be
interesting to study how statistical estimates of the scale of supersymmetry breaking change, once the model is embedded in
string theory.

\bigskip

\noindent
{\bf Acknowledgements}

We thank S.~Thomas for suggesting this problem.
We also thank T. Banks, K.~van den Broek, D.E.~Diaconescu, M.R.~Douglas, J.-F.~Fortin, K.~Intriligator, G.~Moore, S. Ramanujam and especially D. Shih for 
useful suggestions. We would like to thank T. Banks, K.~Intriligator and D. Shih for reading the manuscript.
This research is supported by the Department of Physics and Astronomy at Rutgers University.


\listrefs
\end